\documentclass[a4paper,12pt]{article}
\usepackage{graphicx}
\usepackage{jcappub}
\title{Experimental search for solar hidden photons in the eV energy range using kinetic
mixing with photons}
\author[a]{T.~Mizumoto}
\author[b]{R.~Ohta}
\author[b]{T.~Horie}
\author[b]{J.~Suzuki}
\author[c]{Y.~Inoue}
\author[b,d]{M.~Minowa}

\affiliation[a]{Department of Physics, Graduate School of Science, Kyoto University,\\
Kitashirakawa-Oiwakecho, Sakyo-ku, Kyoto 606-8502, Japan}
\affiliation[b]{Department of Physics, School of Science, The University of Tokyo,\\
7-3-1 Hongo, Bunkyo-ku, Tokyo 113-0033, Japan}
\affiliation[c]{International Center for Elementary Particle Physics, The University of Tokyo,\\
7-3-1 Hongo, Bunkyo-ku, Tokyo 113-0033, Japan}
\affiliation[d]{Research Center for the Early Universe (RESCEU), School of Science, The University of Tokyo,\\
7-3-1 Hongo, Bunkyo-ku, Tokyo 113-0033, Japan}
\emailAdd{minowa@phys.s.u-tokyo.ac.jp}
\abstract{We have searched for solar hidden photons in the eV energy range using a dedicated hidden photon detector. 
The detector consisted of a parabolic mirror with a diameter of 500\,mm and a focal length of 1007\,mm installed in a vacuum chamber, and a photomultiplier tube at its focal point.
The detector was attached to the  Tokyo axion helioscope, Sumico which has a mechanism to track the sun.
From the result of the measurement, we found no evidence for the existence of hidden photons and  set a limit on the photon-hidden photon mixing parameter $\chi$ depending on the hidden photon mass $m_{\gamma '}$. 
}
\begin{document}
\maketitle

\section{Introduction}

A hidden photon is the gauge boson of a hypothetical hidden local U(1) symmetry.
Such symmetries arise in a generic prediction of many extensions of the standard model,
especially in those  based on string theory\cite{Ringwald}.
If the hidden photons are massive but the mass is small enough, the hidden photons have rich phenomenology\cite{Okun1}
at low energy scales.

The dynamics of the
photon-hidden photon system with kinetic mixing is described by the following Lagrangian,
\begin{equation}
\mathcal{L} = -\frac{1}{2}\chi F_{\mu\nu}B^{\mu\nu},
\end{equation}
where $F_{\mu\nu}$ and $B_{\mu\nu}$ represent the ordinary and the hidden photon field, 
respectively\cite{Okun1,Holdom,Foot}. 
When the hidden photon has non-zero mass $m_{\gamma '}$,
it leads to photon-hidden photon oscillations similar
to vacuum neutrino ocillations.
In vacuum, hidden photon $\to$ photon transition probability 
$P_{\gamma' \to \gamma}(\omega)$ is given by:

\begin{eqnarray}
P_{\gamma' \to \gamma}(\omega) = 4\,\chi^{2}\,\sin^{2}\left(\frac{\Delta q\,\ell}{2}\right),
\end{eqnarray}
where $\omega$ is the energy of the photon, 
$\ell$ is the traveling path length 
and $\Delta q$ is the momentum transfer between the photon and hidden photon which is given by:
\begin{eqnarray}\label{eq:}
\Delta q = \omega - \sqrt{\omega^{2} - m_{\gamma'}^{2}} \sim \frac{m_{\gamma'}^{2}}{2\omega},
\end{eqnarray}
assuming $m_{\gamma'} \ll \omega$. 

The matter effects
modify the photon-hidden photon transition probability.
Neglecting photon absorption, it can be written with the effective photon mass $m_{\gamma}$\cite{Popov1,Popov2} as
\begin{equation}
P_{\gamma' \to \gamma}(\omega)  = 
\frac{4\,\chi^{2}\,m_{\gamma'}^{4}}{(m_{\gamma'}^{2} - m_{\gamma}^{2})^{2} 
+ 4\,\chi^{2}\,m_{\gamma'}^{4}} \times \sin^{2} 
\left(\ell \times\frac{\sqrt{(m_{\gamma'}^{2} - m_{\gamma}^{2})^{2} + 4\chi^{2}m_{\gamma'}^{4}}}{4\omega}\right).
\label{eq:prob}
\end{equation}
The effective photon mass $m_{\gamma}$ is defined by the following relation with the momentum $k$
of the photon and the refractive index $n$,
\begin{equation}
m_{\gamma}^2  = \omega^2 - k^2 = -\omega^2 \left(\frac{k^2}{\omega^2} -  1\right) = -\omega^2 (n^2 -  1).
\end{equation}
The probability gets lower as the matter density becomes higher due to the denominator of the oscillation amplitude.

Constraints on the massive hidden photon have been obtained
from precision measurements of Coulomb's law\cite{Okun1,Clmb1,Clmb2}, 
from stellar cooling considerations \cite{Popov1,Popov2,Redondo1}, 
and from the photon regeneration or LSW(Light Shining through Walls) experiments.  
Recently, constraints on $\chi$ for the mass region 
$10^{-4}{\rm eV} < m_{\gamma'} < 10^{-2}{\rm eV}$ have been obtained 
from the results of the ALPS collaboration\cite{LSW1}, the BMV collaboration\cite{LSW2}, the GammeV collaboration\cite{LSW3}, and the LIPSS collaboration\cite{LSW4}. 
More recently, a high energy solar hidden photon search with HP Ge detector\cite{HPGe}  was reported.
Solar axion search experiments are sensitive 
to the keV part of the solar spectrum of hidden photons and the latest CAST results\cite{cast1,cast2} 
have been translated into limits on the photon-hidden photon mixing 
parameter\cite{Redondo1}. 
Bounds on models with additional new particles and a hidden photon 
at a low energy scale could be obtained from astrophysical 
considerations\cite{astroph2,astroph3,astroph4}.
If hidden photons exist, production of them leads to distortions in the cosmic microwave background(CMB) spectrum.
The CMB spectrum data provided by the Far Infrared Absolute
Spectrophotometer (FIRAS) on board of the COBE constrained the hidden
photon existence\cite{CMB}.

Since hidden photons can be produced through mixing with ordinary photons, 
the sun could be a source of low energy hidden photons. 
The coherence length of the photon-hidden photon 
oscillations is much shorter than the distance from the sun to the earth.
Therefore, the transition probability is 
$2\chi^{2}$ and the flux of hidden photons from the solar surface\cite{Redondo1} 
is calculated to be
\begin{eqnarray}
\frac{{\rm d}\Phi^{s}}{{\rm d}\omega} \simeq \chi^{2}\,(4.2\,\times10^{18})\,\frac{\omega^{2}}{e^{\omega/T_{0}}\,-\,1}\,\frac{1}{\rm{eV}^{3}\,\rm{cm^{2}}\,\rm{s}}.
\label{surface}
\end{eqnarray}

In addition to the above flux, much higher flux of hidden photons is expected 
from photon-hidden photon oscillations in the bulk solar interior
with a higher emitting volume and a higher temperature. 
For $m_{\gamma'}$ well below the eV, one can use the following conservative estimate 
for the bulk component of the hidden photon flux 
at the earth\cite{Redondo1}:

\begin{eqnarray}
\frac{{\rm d}\Phi^{b}}{{\rm d}\omega} \simeq \chi^{2}\,\left(\frac{m_{\gamma'}}{\rm{eV}}\right)^{4}\,10^{32}\frac{1}{\rm{eV}\,\rm{cm}^{2}\,\rm{s}} \qquad \rm{for} \quad \omega = \mbox{1--5}\,\rm{eV},
\label{bulk}
\end{eqnarray}
which exceeds the surface contribution except for masses $m_{\gamma'} \leq 10^{-4}$ eV. 

Recently, more refined estimation of the bulk flux is given by the same author\cite{Redondo2,Redondo3}
taking into account the resonant production of hidden photons in a thin spherical solar shell,
where the effective photon mass $m_\gamma$ is equal to the hidden photon mass $m_{\gamma '}$.
They claim that the resonant production dominates over the emission from the rest of the sun
and gave hidden photon flux estimation for four typical cases of the hidden photon mass $m_{\gamma '}$ = 0, 0.01, 0.1 and $1\, $eV. 

In this paper, we report on the direct experimental search for the flux of solar hidden photons.

\begin{figure}
 \centering
 \includegraphics[width=9cm]{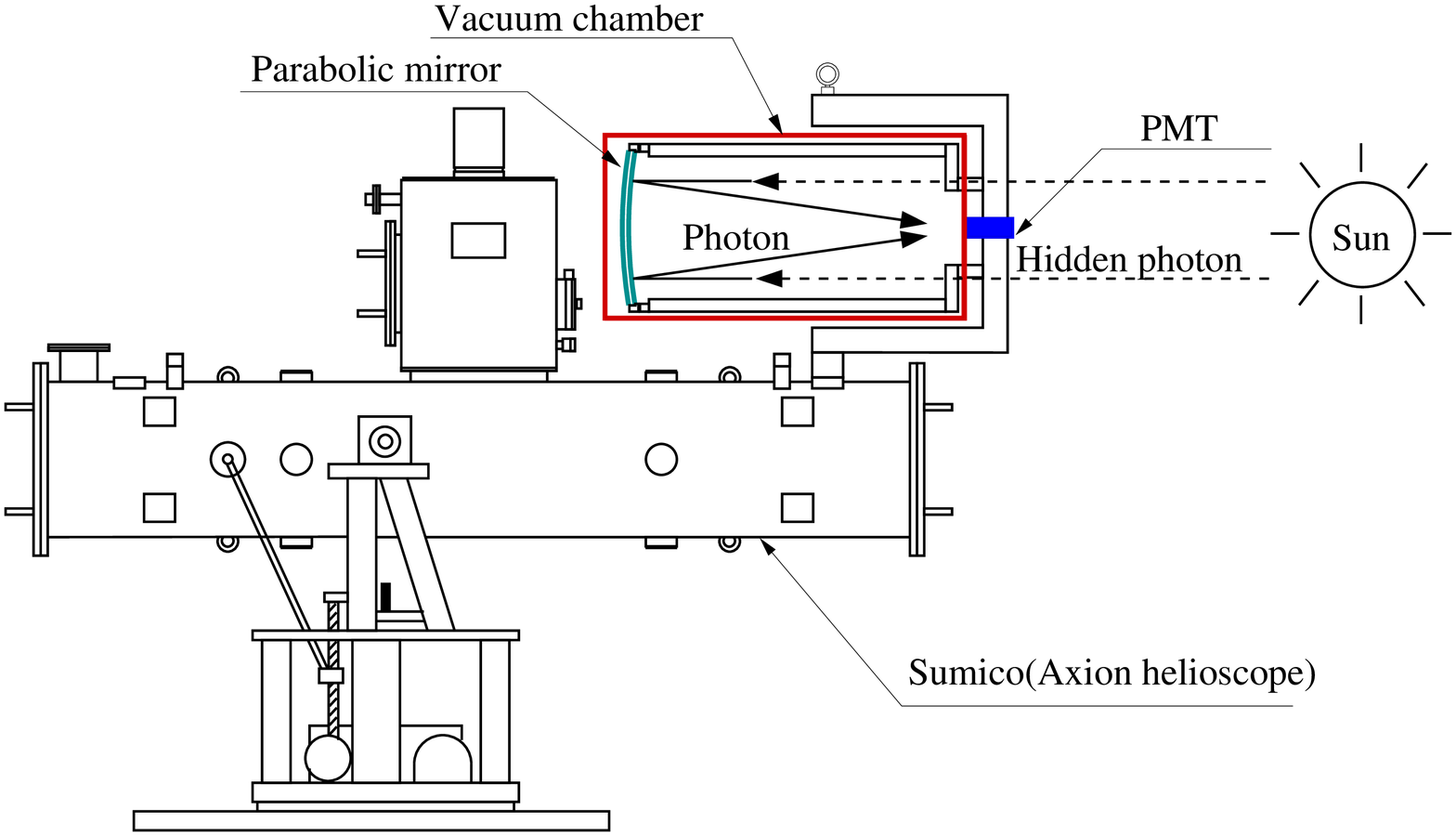}
 \caption{The schematic view of Sumico (Tokyo Axion Helioscope) and the solar hidden photon detector.}
 \label{apparatus}
\end{figure}

\section{Experimental apparatus}

For this experiment, we built a dedicated solar hidden photon 
detector and mounted it on Sumico, the Tokyo axion helioscope\cite{Sumico, Sumico1, Sumico2, Sumico3}
as shown in fig.~\ref{apparatus}. 
Sumico has an altazimuth mount with a driving range from $-28^\circ$ to $28^\circ$
in altitudinal direction and $360^\circ$ in azimuth.
Sumico is designed to search for the solar axion which is a particle introduced 
to solve the strong CP problem\cite{PQ1,PQ2}.

The overall tracking accuracy is better than 0.5\,mrad both in altitudinal and
azimuthal direction. 
Main components of the errors are a fluctuation of the
turntable and a possible misalignment of the magnet aperture and the helioscope axis.
The guidance of the helioscope movement is provided by the tracking software. 
In order to calculate the position of the sun, the U. S. Naval Observatory Vector 
Astronomy Subroutines (NOVAS-C ver 2.0.1)\cite{NOVAS-C} is used. 
NOVAS-C calculates the topocentric position of the sun with less than 2 arcseconds
(9.7 $\mu$rad) error.
The altitudinal origin was determined from a spirit level. 
While the sun is not directly visible from the laboratory in the basement floor, 
the azimuthal origin was first determined by a gyrocompass, 
which detects the north direction by the rotation of the earth within an
error of 8 arcseconds(39 $\mu$rad), and then it was introduced to the laboratory with a theodolite.
The overall tracking error is negligible in our measurements.

The solar hidden photon 
detector consists of a vacuum chamber, a parabolic mirror 
and a single photon detector.

The vacuum chamber holds the conversion region in vacuum 
to keep the hidden photon $\to$ photon conversion probability high enough. 
It is a cylinder made of 1.5-mm thick stainless steel plates 
with wrinkles on its side for the mechanical reinforcement. 
The inner diameter of the cylindrical vacuum chamber is 567mm and its length 
is 1200mm. 
The cylinder can be divided into two parts: 
the cylinder main body 
with a closed bottom on one end and its lid on the other end. 
The lid is equipped with a vacuum gauge and an electronic thermometer. 
An ISO KF40 port is placed at the center of the lid and stands up toward the inside. 
The port is used to install a photomultiplier tube (PMT).
A quartz glass vacuum window is attached to the port to transmit conversion photons 
to the PMT which is set at the atmospheric pressure side. 

A parabolic mirror was used to collect the conversion photons to the PMT at its focal point.
The mirror is made of soda glass and aluminium is deposited on the surface 
of the mirror. 
The mirror is 500\,mm in diameter, 19\,mm thick, 1007\,mm focal length and the focal spot diameter is 1.5\,mm. 
The mirror is mounted on a mirror holder made of two aluminium rings.
One holds the mirror directly by four clamps with silicon rubber pads.
Another part of the mirror holder is attached to four channel steels 
which hold the parabolic mirror and its holder. 
Both are fixed to each other with three pairs of pushing and pulling
 bolts at the corners of an equilateral triangle, with which we can
 adjust the mirror axis so that the center of the PMT is on the optical
 axis.
The reflectance of the parabolic mirror is measured by the manufacturer
as a function of the wavelength of the photon. 
It is higher than 80\% over the range between 300 and 650\,nm
with the maximum at around 400\,nm.

We used a photon counting PMT as a detector of the photon 
which is generated in the process of hidden photon $\to$ photon conversion. 
We selected a PMT, Hamamatsu Photonics R3550P because of its low dark count rate. 
It is a head-on type PMT and the tube size is 25\,mm in diameter. 
It has a low noise bialkali photocathode whose effective area is 22\,mm 
in diameter and it is sensitive to photons of wavelength range 
300--650\,nm with a peak quantum efficiency of 17\,\%. 

Single and multi photon events detected by the PMT make  current pulses 
which enter a charge-sensitive preamplifier (ORTEC 113) 
and a shaping amplifier (ORTEC 572).
The signal is then sent to an ADC (Laboratory Equipment Corp. 2201A) and the multichannel analyser\,(MCA) spectrum is taken and recorded by a PC every 100$\,$s live time.

The inner pressure of the vacuum chamber was measured by a vacuum gauge (Balzers PKR250).
The temperatures of the PMT and the vacuum chamber were measured by Pt100 thermometers
and recorded by another PC.
During the solar tracking and background measurements, 
the inner pressure of the vacuum chamber was lower than
$(5 \pm 2) \times 10^{-3}\,$Pa at a 
temperature of 23$^{\circ}$C.
The effect of this remaining gas on the conversion probability 
equation~(\ref{eq:prob})
is negligible.

\section{Measurement and analysis}
If a hidden photon is converted into a photon in the vacuum chamber,
it would be detected by the PMT as a single photon event.
Before starting the measurement, 
the shape of a single photon spectrum in the MCA was measured
by illuminating the PMT
with a blue LED with sufficiently low current pulses.
It was fitted by a gaussian function and was later used as the template for the single photon analysis of the hidden photon search.

The solar tracking measurements were done around the time of sunrise and sunset
with tracking time of about 5 hours each.
Background measurements were also done before and after the solar tracking measurement
by directing the detector away from the sun.
All the measurements were done from October 26, 2010 till November 16, 2010 (22days). 

To find out the possible evidence of solar hidden photons from the data of 
the measurement, we subtract the background spectrum from the solar tracking spectrum. 
We must eliminate some systematic effects which have nothing to do with 
the solar hidden photons. 

It is well known that the dark count rate gets lower as time passes after an operating voltage is applied.
We, therefore, waited for four days until the time dependence on the dark count rate became negligible. 

\begin{figure}
\begin{center}
  \includegraphics[width=14cm]{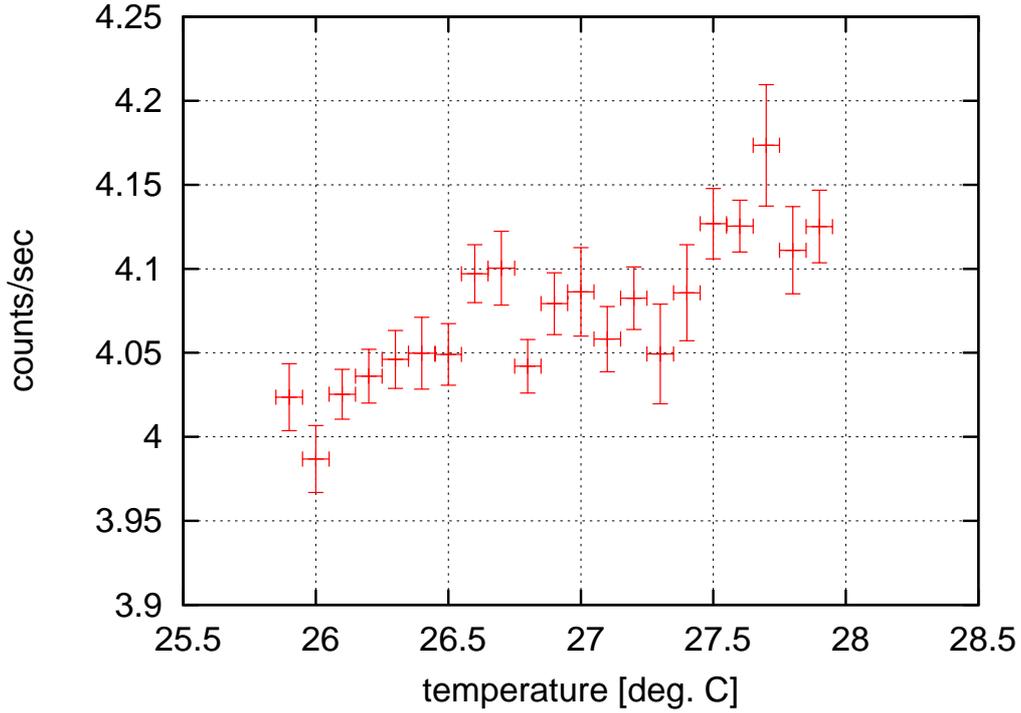}
\end{center}
\caption{Temperature dependence of the dark count rate.}
\label{fig:temp-dep}
\end{figure}

\begin{figure}
\begin{center}
  \includegraphics[width=9cm]{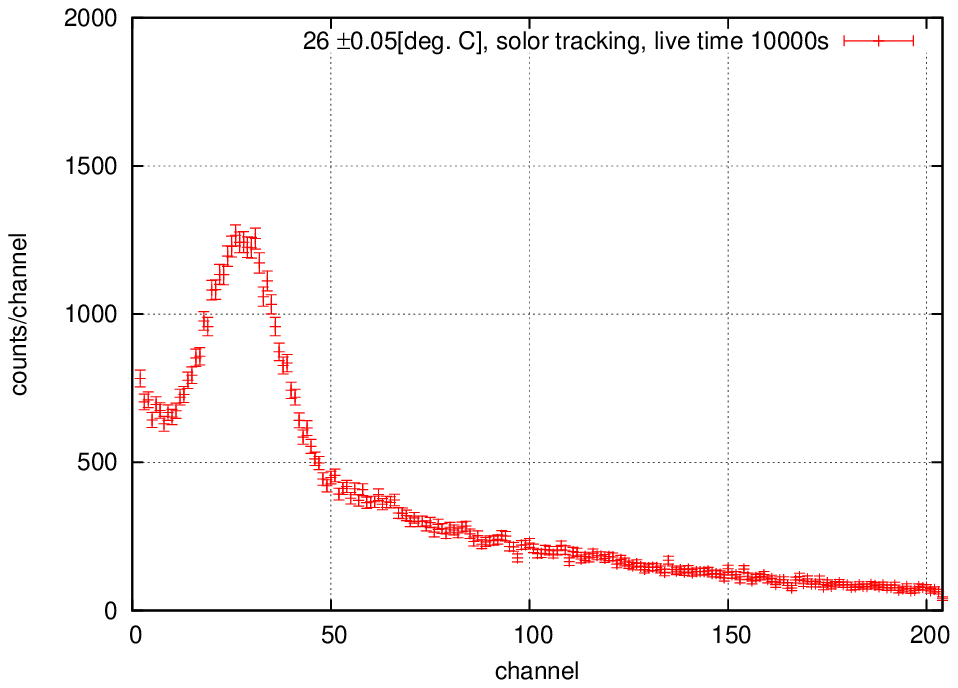}
  \includegraphics[width=9cm]{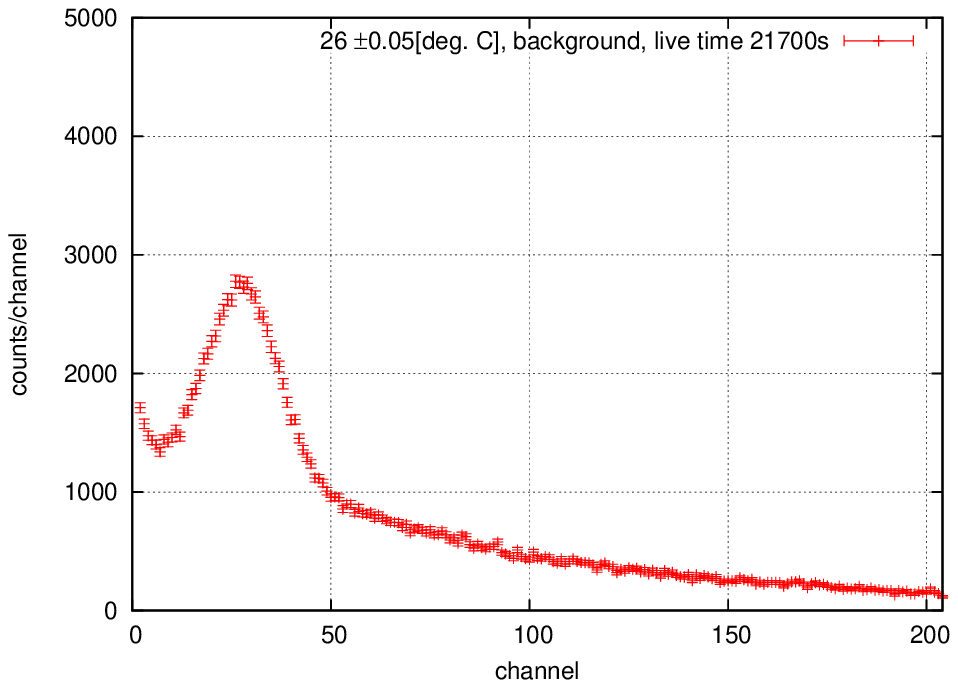}
  \includegraphics[width=9cm]{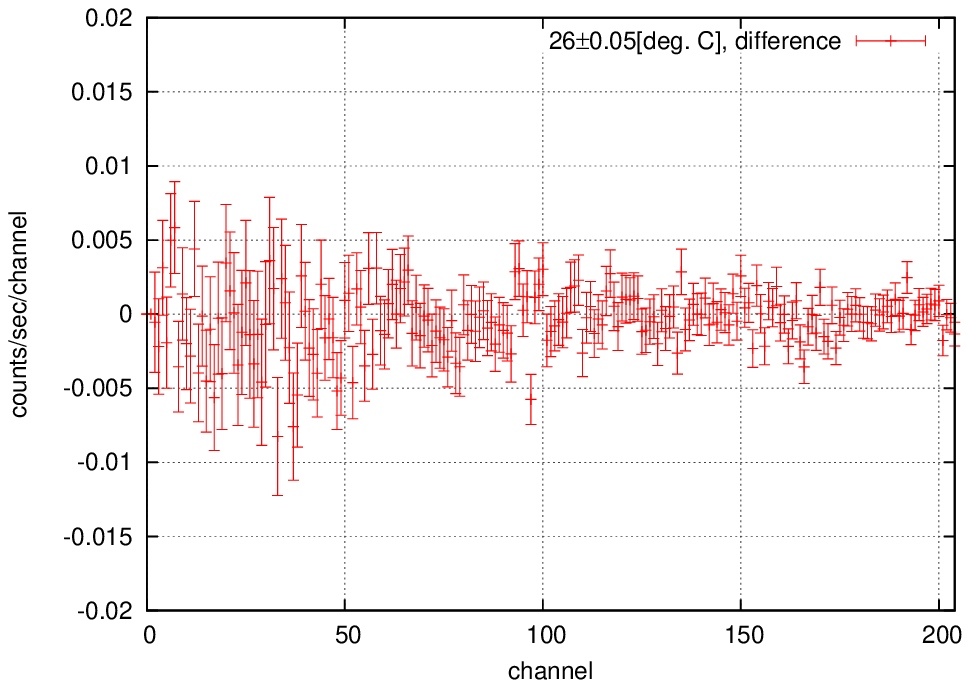}
\end{center}
\caption{Spectrum examples (PMT temperature 26.0 $\pm$ 0.05 $^{\circ}$C). Top: solar tracking spectrum. 
Middle: background spectrum. 
Bottom: spectrum after background subtraction.}
\label{fig:graph_no2}
\end{figure}

Next, a temperature dependence of the dark count rate might cause 
a systematic effect on the background subtraction.
Fig.$\,$\ref{fig:temp-dep} shows observed temperature dependence of the dark count rate.
To avoid the effect, we subtracted background isothermally.
First, we grouped the solar tracking- and background-spectra
each with 100\,s of live time into 
21
temperature bins of 0.1$^\circ$C interval
each whose central values ranging from 25.9 to 27.9$^\circ$C.
Then, 
we apply the background subtraction in every temperature bin and 
obtained 21 residual spectra.
%
Fig.$\,$\ref{fig:graph_no2} shows the solar tracking data, 
background data and residual spectrum in the temperature bin $26 \pm 0.05 \,^{\circ}$C, as an example. 
The former two show peaks of single photoelectron events.
Finally, we combined the residual spectra of all the temperature bins 
%
taking into account errors in them,
and obtained the total residual spectrum.

In the above procedure, we took only the data during the holidays
when the air conditioning system was switched off,
because we observed abrupt room temperature changes on weekdays 
due to automatic switching of the air conditioning system of the building.
The subtraction scheme might fail 
when temperature changes so quickly that the thermometer does not 
follow the PMT temperature.

\begin{figure}
\begin{center}
  \includegraphics[width=14cm]{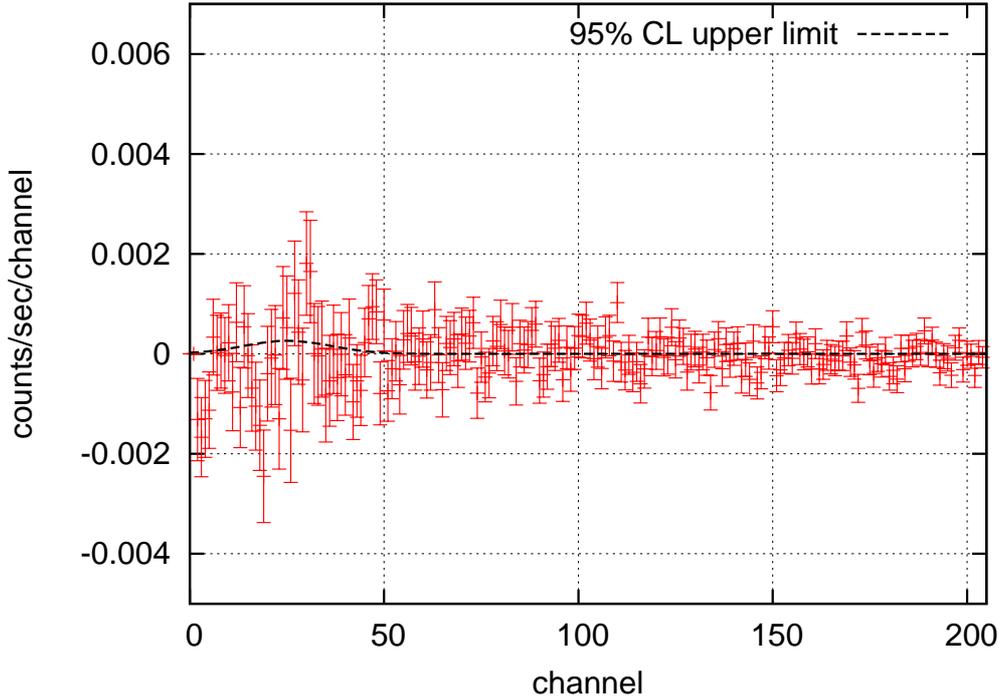}
\end{center}
\caption{Total residual spectrum and the 95\% confidence level upper limit.}
\label{fig:gnuplot_final_fit}
\end{figure}

The final result is shown in fig.$\,$\ref{fig:gnuplot_final_fit}.
We then estimated how many single photons could there be in the total residual spectrum
by fitting the magnitude of the gaussian template function to it.
The best fit was obtained with
\begin{equation}
N_{\rm fit} = (-7.9\pm 6.5({\rm stat.})\pm 3.4({\rm sys.}))
\times 10^{-3}[{\rm s}^{-1}].
\end{equation}
The systematic errors considered include an effect of Cherenkov light emitted in the quartz glass
vacuum window and the PMT window by cosmic muons.
They might come into the PMT and be observed as a single photon event.
Since cosmic muons have directional dependence, background subtraction might fail to give a fake effect.
The finite bin width of the temperature might cause a systematic effect as well.
We would like to make the bin width as narrow as possible, 
but too narrow a binning is impractical.
The systematic error is estimated with the adopted temperature bin width and the temperature dependence of the dark count rate estimated in fig.$\,$\ref{fig:temp-dep}.
The third thing to be considered is drift effect of the dark noise rate.
As already described in the above paragraph, we waited for four days to start the measurement
until the dark count rate got stable.
A residual drift of the dark count rate after four days could give a systematic error.
Possible gain drift of the PMT was monitored by tracking the single photoelectron peak of the spectrum.
It amounted to $\pm 2$\% during the measurement and could cause only negligible systematic error. 
All these systematic errors and the statistical error are estimated and summarized 
in table~\ref{tab:syst-stat}.

The 95\% confidence level upper limit to the hidden photon counting rate was estimated 
from the fitting taking the statistical and systematic errors into account;

\begin{equation}
N_{\rm UL95} = 1.02\times 10^{-2}{\rm s}^{-1} .
\label{obs}
\end{equation}

\begin{table}
\begin{center}
\caption{Systematic and statistical errors}
\begin{tabular}{lr}
\hline
item & value(counts/s)\\
\hline
muonic Cherenkov light & 1.1$\times 10^{-3}$\\
temperature bin & 2.8$\times 10^{-3}$\\
dark noise drift & 1.5$\times 10^{-3}$\\
statistical & 6.5$\times 10^{-3}$\\
%
%
\hline
total & 7.4$\times 10^{-3}$\\
\hline
\end{tabular}
\label{tab:syst-stat}
\end{center}
\end{table}

The obtained upper limit $N_{\rm UL95}$ is now compared with the count rate $N_{\rm exp}$ expected by
the hidden photon model with given parameters;

\begin{eqnarray}
N_{\rm exp} = \int {\rm d}\omega \frac{{\rm d}\Phi}{{\rm d}\omega} (\chi, m_{\gamma '}, \omega) & \times&
\eta_{\rm mirror} (\omega) \eta_{\rm window} \eta_{\rm PMT} (\omega)\nonumber \\ 
& \times & S P_{\gamma ' \to \gamma}(\chi, m_{\gamma '}, \omega, n(p, T, \omega), \ell),
\label{exp}
\end{eqnarray}
where $\frac{{\rm d}\Phi}{{\rm d}\omega}$ is the solar hidden photon spectral flux 
at the surface of the earth, 
$\eta_{\rm mirror}$
is the reflectance of the parabolic mirror, 
$\eta_{\rm window}$ is the transmittance of
the quartz glass window, 
$\eta_{\rm PMT}$ is the detection efficiency of the PMT, 
$S$ is the area of the conversion region of the experimental apparatus, 
$P_{\gamma ' \to \gamma}$ is the hidden photon to photon conversion probability in the apparatus
and $n$ is the refractive index of the conversion region as a function of 
the pressure $p$ and the temperature $T$. 

From equations (\ref{obs}) and (\ref{exp}), the upper limit to the mixing angle $\chi$
as a function of the hidden photon mass $m_{\gamma '}$ is calculated.
In the calculation, possible systematic errors in the parameters of equation (\ref{exp}) have been 
taken into account; 
$\eta_{\rm mirror}$, $\eta_{\rm window}$, $\eta_{\rm PMT}$, $S$ and $\ell$. 
They are relatively small compared to the systematic errors of $N_{\rm fit}$.

For the solar hidden photon flux $\frac{{\rm d}\Phi}{{\rm d}\omega}$ in equation(\ref{exp}),
we assumed two cases.
One is the sum of the conservative estimations\cite{Redondo1}, equations(\ref{surface})  
and (\ref{bulk}).
The other is the more refined flux calculation\cite{Redondo2, Redondo3} 
with the resonant hidden photon production in the spherical solar shell beneath the surface.
 
\begin{figure}
\begin{center}
  \includegraphics[width=15cm]{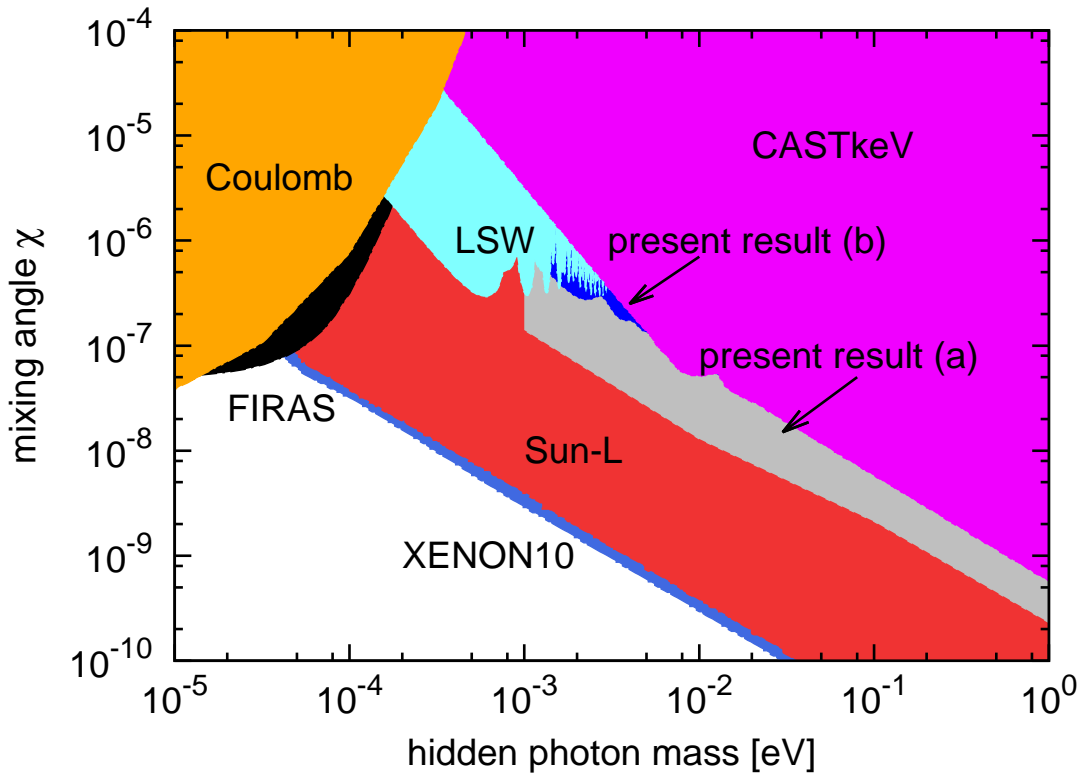}
\end{center}
\caption{95\,\% Confidence level upper limits to the mixing angle $\chi$ set by this experiment.
Present result (a) is obtained on the basis of the newer refined solar hidden photon flux calculation\cite{Redondo2, Redondo3}
and (b) on the basis of the older conservative estimations\cite{Redondo1}, equations(\ref{surface})  
and (\ref{bulk}). 
Filled areas are excluded by other experiments.
The regions excluded by  precision measurements of Coulomb's law\cite{Clmb1, Clmb2},
LSW(Light Shining through Walls) experiments\cite{LSW1,LSW2,LSW3,LSW4},
the CAST experiment\cite{Redondo1} and FIRAS CMB spectrum\cite{CMB}
are marked ``Coulomb", ``LSW", ``CAST keV" and ``FIRAS", respectively.
The solar luminosity constraints\cite{An1, Redondo4} in the longitudinal channel 
and XENON10 limits\cite{An2} on the longitudinal solar hidden photons
are marked ``Sun-L" and ``XENON10", respectively.
}
\label{fig:gnuplot_result}
\end{figure}

Thus obtained 95\% confidence level upper limit to the mixing angle $\chi$ is shown in fig. \ref{fig:gnuplot_result}.
We also show the limits set by other experiments  with filled areas.
The regions excluded by  precision measurements of Coulomb's law\cite{Clmb1, Clmb2},
LSW(Light Shining through Walls) experiments\cite{LSW1,LSW2,LSW3,LSW4} and
the CAST experiment\cite{Redondo1}
are marked ``Coulomb", ``LSW" and ``CAST keV", respectively.
The excluded region by FIRAS CMB spectrum\cite{CMB} is marked ``FIRAS".

After submitting the original manuscript, the authors have learned of the works\cite{An1, Redondo4}  reporting 
the calculations of hidden photon emission from the sun and other stars. 
They emphasized the effect of longitudinal-mode hidden photon emission 
underestimated in the previous calculations\cite{Redondo1, Redondo2, Redondo3},
and gave the solar luminosity constraints on the hidden photon parameters, 
which we added in fig. \ref{fig:gnuplot_result}.
Another work\cite{An2} reanalyzed the published data of XENON10 dark matter search experiment
and compared them with the predominant longitudinal-mode solar hidden photon flux to obtain
the constraints on the hidden photon parameters.
The constraints is also added in fig. \ref{fig:gnuplot_result}.

Even with these new works, on the other hand,
the present result stays unchanged because it solely rely on 
the transverse-mode solar hidden photon emission flux\cite{Redondo1, Redondo2, Redondo3},
which is valid as far as the transverse-mode is concerned\cite{An1, Redondop}.

\section{Conclusion}
We have searched for solar hidden photons in the eV energy range using a dedicated detector for the first time. 
The detector was attached to the  Tokyo axion helioscope, Sumico which has a mechanism to track the sun.
From the result of the measurement, there is no evidence of the existence of 
hidden photons and we set a limit on photon-hidden photon mixing parameter $\chi$ 
depending on the hidden photon mass $m_{\gamma '}$. 
The present result improved the existing limits given by the LSW
experiments and the CAST experiment in the hidden photon mass
region between $10^{-3}$ and 1\,eV.
With recent new calculations of the longitudinal-mode hidden photon,
more stringent limits came out by the solar luminosity consideration and also by the reanalysis of XENON10 data,
while the present result is based on the search for the transverse-mode solar hidden photons.

\section*{Acknowledgements}
The authors thank all the historical members of Sumico experiment because the present hidden photon search
solely owes them for her altazimuth tracking system.
T.~Horie acknowledges support by Advanced Leading Graduate Course for Photon Science (ALPS) 
at the University of Tokyo.
T.~Mizumoto would like to thank support by
the Grant-in-Aid for JSPS Fellows Grant Number 229033 and by the Global COE Program ``the Physical
Sciences Frontier", MEXT, Japan.
R.~Ohta would like to thank support by the Grant-in-Aid for JSPS Fellows Grant
Number 207600.
This reaserch is supported by the Grant-in-Aid for challenging Exploratory Research
by MEXT, Japan,
and by the Research Center for the Early Universe, School of Science, the University of Tokyo.
The authors acknowledge useful communication with J. Redondo on the solar hidden photon flux.

\end{document}